\documentstyle[twocolumn]{article}
\title{String baryon model ``triangle":\\
hypocycloidal solutions}
\author{G.S. Sharov\\
\small\it Tver State University, Zhelyabova 33, 170000, Tver, Russia\\
\parbox{14.2cm}{\small\rm\parindent=0.5cm\medskip
The considered model of baryon consists of
three pointlike masses (quarks) bounded pairwise by relativistic
strings forming a curvilinear triangle. Classic analytic solutions
for this model corresponding to a planar uniform rotation
about the system center of mass are found and
investigated. These solutions describe a rotating curve
composed of segments of a hypocycloid. The curve is a curvilinear
triangle or --- a more complicated configuration
with a set of internal massless points moving at
the speed of light. Different topological types of these
motions are classified in connection with different forms of
hypocycloids in zero quark mass limit.
An application of these solutions to description of baryon states
on the Regge trajectories is considered.}}
\date{}

\begin{document}
\maketitle

\smallskip

\centerline{\bf INTRODUCTION}

\medskip

The string baryon model ``triangle" is genetically con\-nec\-ted with
the meson model of relativistic string with massive ends [1,2].
The latter model including two pointlike massive quarks bounded by
a relativistic string gives possibility to describe the meson
orbitally excited states on the Regge trajectories [3].

On a base of this meson model string models of baryon were suggested
in some variants [4--7]. These variants differ from each other
in the type of spatial junction of three pointlike quarks by
relativistic strings: a) the first quark is bounded with the second
and the second quark --- with the third; b) the ``three-string"
model or Y-configuration
with three strings from three quarks joined in the fourth massless
point; c) the quark-diquark model and d) the ``triangle" model.
The first variant was investigated qualitatively [4],
the ``three-string" [4\,--\,6] and meson-like quark-diquark
models [7,8] --- in a more detailed way.

In the present paper the ``triangle" model of baryon [9] is under
consideration.
In this model three material points (quarks) are pairwise connected
by three relativistic strings forming a curvilinear triangle in
space at each instant of time. If tensions of these three strings are
equal, such an object could be regarded as a closed string carrying
three pointlike masses. From the point of view of describing
quark strong interaction in the orbitally excited baryon this model
looks rather natural in comparison with three others.
Some arguments in favour of the ``triangle" baryon model in
comparison with Y-configuration are given by Cornwall [10]
in the QSD Wilson loop operator approach.

Transformation of the ``meson" string with massive ends or the
three-string model of baryon to the model ``triangle" results in
some additional difficulties. In particular, a string world surface
in this model has discontinuities of derivatives on quark
trajectories; a parametrization with this trajectories as
co-ordinate curves doesn't exist in general; space-like
co-ordinate lines are not closed in general.

In the present paper these difficulties were overcome and classic
analytic solutions were found for a set of motions --- uniform
planar rotations of the system. This kind of motions is an analog
and generalization of well known rotations of a straight relativistic
string with massive ends [1\,--\,3]. The latter class of motions was a
base of applying this model [3,8] and the relativistic tube
model [11] to description the meson Regge trajectories.

In this paper rotational motions in the baryonic model ``triangle"
and their applications are investigated. In Sec.~1 equations of
evolution and conditions on the quark trajectories are deduced
from the action of the system. In Sec.~2 solutions of these
equations corresponding to rotational motions of the system
(quarks and the string of hypocycloidal form) are described and
classified. In Sec.~3 possibility of description the baryon
states on the Regge trajectories by these solutions is discussed.

The string solutions obtained here in Sec.~2 are applicable
not only to the particle physics, but to various branches of
string or M-brane theory. In particular, the massive points placed
on the string (the number of these points is arbitrary)
could be regarded as 0-branes.

Note that the rotational solutions of the considered type also take
place for a closed massless string. Such a string has a form of a
rotating hypocycloid with singular points moving at the speed of light.

\bigskip

\centerline{\bf I. MODEL AND EQUATIONS}

\medskip

Let us consider the baryon model ``triangle" as a closed
relativistic string with tension $\gamma$ carrying three point\-like
masses $m_1$, $m_2$, $m_3$. The action of this system is [9]
\begin{equation}
S=-\int\limits_{\tau_1}^{\tau_2}\!\biggl\{\gamma\!
\int\limits_{\sigma_0(\tau)}^{\sigma_3(\tau)}\!\!\sqrt{-g}\,d\sigma+
\sum_{i=1}^3m_i\sqrt{V_i^2(\tau)}\biggr\}d\tau.
\end{equation}
Here $g=\dot X^2X'{}^2-(\dot XX')^2$ is a determinant of indu\-ced
metric on a string world surface
$\,x^\mu=X^\mu(\tau,\sigma)$, $\,\mu=0,1,\dots$ in $d$-dimensional
Minkowski space with signature $+,-,-,\dots$;
$\,\dot X^\mu=\partial_\tau X^\mu$,
$X'{}^\mu=\partial_\sigma X^\mu$,
the speed of light in these units $c=1$,
$\;(\tau,\sigma)\in\Delta=\Delta_1\cup\Delta_2\cup\Delta_3$,\linebreak
$V_i^\mu=\frac d{d\tau}X^\mu(\tau,\sigma_i(\tau))$ is a tangent
vector to the i-th quark trajectory with an inner equation
$\sigma=\sigma_i(\tau)$, $i=0,\,1,\,2,\,3$ (Fig.~1).
\begin{figure}[th]
\begin{picture}(86,40)
\put(5,2){\vector(1,0){79}} \put(25,0){\vector(0,1){39}}
\thicklines
\put(9,5){\line(1,0){67}} \put(6,35){\line(1,0){75}}
\put(9,5){\line(1,6){1}} \put(10,11){\line(0,1){4}}
\put(10,15){\line(-1,6){1}} \put(9,21){\line(-1,5){1}}
\put(8,26){\line(-1,4){1}} \put(7,30){\line(-1,5){1}}
\put(28,5){\line(1,5){1}} \put(29,10){\line(1,6){1}}
\put(30,16){\line(0,1){4}} \put(30,20){\line(1,6){1}}
\put(31,26){\line(1,5){1}} \put(32,31){\line(1,4){1}}
\put(50,5){\line(1,4){1}} \put(51,9){\line(1,5){1}}
\put(52,14){\line(1,6){1}} \put(53,20){\line(0,1){3}}
\put(53,23){\line(-1,6){1}} \put(52,29){\line(0,1){2}}
\put(52,31){\line(1,4){1}}
\put(76,5){\line(1,6){1}} \put(77,11){\line(0,1){5}}
\put(77,16){\line(1,6){1}} \put(78,22){\line(1,5){1}}
\put(79,27){\line(1,4){2}}
\put(81,3){$\sigma$} \put(26,38){$\tau$}
\put(20,36){$\tau_2$} \put(20,6){$\tau_1$}
\put(0,12){$\sigma_0(\tau)$}
\put(30.5,12){$\sigma_1(\tau)$} \put(53,12){$\sigma_2(\tau)$}
\put(78,14){$\sigma_3(\tau)$} \put(15,23){$\Delta_1$}
\put(39,23){$\Delta_2$} \put(62,23){$\Delta_3$}
\end{picture}
\caption{}
\end{figure}
The equations $\sigma=\sigma_0(\tau)$ and $\sigma=\sigma_3(\tau)$
define the trajectory of the same --- 3-rd quark.
It is connected with the fact that the string is closed and may
be rewritten in the common form
\begin{equation}
X^\mu(\tau,\sigma_0(\tau))=X^\mu(\tau^*,\sigma_3(\tau^*)).
\end{equation}

Note that the parameters $\tau$ and $\tau^*$ in these two
parametrizations of one curve (2) aren't equal in general.
It means
that co-ordinate curves $\tau=\,$const on the world surface are
not closed --- the beginning of this curve at $\sigma=\sigma_0$
doesn't coincide spatially with its end at $\sigma=\sigma_3$.

The equality $\tau=\tau^*$ may be obtained only by a special choice of
$\tau$ and $\sigma$, for example, $\tau=t\equiv X^0$.

The parametrization of the world surface $X^\mu(\tau,\sigma)$
is continuous in $\Delta$, but on the lines $\sigma_i(\tau)$ its
derivatives (except for tangential $V_i^\mu$ and
$\frac d{d\tau}V_i^\mu$) have discontinuities in general.
Nevertheless, by a local choice of parameters $\tau$ and $\sigma$
we can obtain the induced metric
$ds^2=\dot X^2d\tau^2+2(\dot XX')\,d\tau\,d\sigma+X'{}^2d\sigma^2$
continuous on these lines.
The action (1) is invariant with respect to an arbitrary
non-degenerate reparametrization
$\tau=\tau(\tilde\tau,\tilde\sigma)$,
$\sigma=\sigma(\tilde\tau,\tilde\sigma)$.

The equations of motion and the boundary conditions on the quark
trajectories in this model are deduced by variation and minimization
of action (1). This procedure is partially similar to that for
the model of relativistic string with massive ends [2] and results
in the same equations of motion
\begin{equation}
\frac\partial{\partial\tau}\frac{\partial L}{\partial\dot X^\mu}
+\frac\partial{\partial\sigma}\frac{\partial L}{\partial X'{}^\mu}=0,
\quad(\tau,\sigma)\in\Delta, \quad L\equiv\sqrt{-g}.
\end{equation}
But to derive
boundary conditions in the model ``triangle" we are to take into
account the discontinuities of $\dot X^\mu$, $X'{}^\mu$ on the lines
$\sigma=\sigma_i(\tau)$. Thereby the term
$\int\!\!\int_\Delta\bigl[\frac\partial{\partial\tau}\bigl
(\frac{\partial L}{\partial\dot X^\mu}\delta X^\mu\bigr)+
\frac\partial{\partial\sigma}\bigl(\frac{\partial L}
{\partial X'{}^\mu}\delta X^\mu\bigr)\bigr]\,d\tau\,d\sigma$
in $\delta S[X^\mu]$ trans\-formed using the Green's formula equals
the sum of three curvilinear integrals of internal boundary values
along the borders of the domains $\Delta_1$, $\Delta_2$, $\Delta_3$
and therefore --- in the following boundary conditions:
\begin{eqnarray}
m_i\frac d{d\tau}\,\frac{V_{i\mu}}{|V_i|}-\gamma\Bigl[\frac
{\partial L}{\partial X'{}^\mu}-\frac{\partial L}{\partial\dot X^\mu}
\sigma_i'(\tau)\Bigr]\Bigl|_{\sigma=\sigma_i+0}\;\nonumber\\
+\gamma\Bigl[\frac{\partial\stackrel{\phantom{-}}{L}}
{\partial X'{}^\mu}-\frac{\partial L}{\partial\dot X^\mu}
\sigma_i'\Bigr]\Bigr|_{\sigma=\sigma_i-0}=0,\quad i=1,2,3.
\end{eqnarray}
For the third quark ($i=3$) in the first two summands we are to put
$\sigma=\sigma_0(\tau)$, and in the third we put $\sigma=
\sigma_3(\tau^*)$ in accordance with the closure condition (2).

From the physical point of view equations (4) are the 2-nd
Newtonian law for the material points $m_i$, moduli of the applied
tension forces equal to $\gamma$.

Let the induced metric on the world surface be con\-formally flat,
i.e., conditions of orthonormality be tied:
\begin{equation}
\dot X^2+X'{}^2=0,\qquad(\dot XX')=0.\end{equation}

These equalities in $\Delta$ may always be obtained
by the reparametrization $\tau=\tau(\tilde\tau,\tilde\sigma)$,
$\sigma=\sigma(\tilde\tau,\tilde\sigma)$
(new co-ordinate lines
$\tilde\tau\pm\tilde\sigma={}$const on the world sheet are integral
curves of equations
$\dot X^2\,d\tau+\big[(\dot XX')\pm L\big]\,d\sigma=0$).
We use the same notation for $\tau$ and $\sigma$ below and suppose
that equalities (5) are satisfied.

Under conditions (5)
the equations of motion (3) become linear
\begin{equation}\ddot X^\mu-X''{}^\mu=0,\end{equation}

Eqs.~(5) and (6) are invariant with respect to repara\-metrizations
$\tau\pm\sigma=f_\pm(\tilde\tau\pm\tilde\sigma)$ [2]. Choosing
these two arbitrary functions $f_\pm$ one can fix two (of four)
functions $\sigma_i(\tau)$, for example, in the form
\begin{equation}
\sigma_1(\tau)=0,\qquad\sigma_2(\tau)=\pi.\end{equation}

The boundary equations (4) on these lines under conditions (5)
and (7) take the form
\begin{eqnarray}
m_i\frac d{d\tau}\,\frac{\dot X^\mu(\tau,\sigma_i)}
{\bigl[\dot X^2(\tau,\sigma_i)\bigr]^{1/2}}
-\gamma X'{}^\mu(\tau,\sigma_i+0)\nonumber\\
{}+\gamma X'{}^\mu(\tau,\sigma_i-0)=0,\qquad i=1,2.
\end{eqnarray}

Reparametrizations of the mentioned type with
$f_+(\eta)=f_-(\eta)=\eta+\phi(\eta)$, $\phi(\eta+2\pi)=\phi(\eta)$,
$|\phi'(\eta)|<1$ [12] preserving the form of equations (7) don't
permit to fix $\sigma_3={}$const (or $\sigma_0={}$const)
for all $\tau$ in general.

Thus choosing $\tau$ and $\sigma$ one can't fix three functions
$\sigma_0(\tau)$, $\sigma_3(\tau)$ and $\tau^*(\tau)$ for an
arbitrary motion in a convenient form. The necessity of determinating
these functions from initial data essentially sophisticates the
initial-boundary-value problem for the model ``triangle" in
comparison with the string model of meson [13].

In the present paper the functions $\sigma_0(\tau)$,
$\sigma_3(\tau)$ and $\tau^*(\tau)$ are defined from properties of
symmetry for a class of uniform planar rotations of the system.

\bigskip

\centerline{\bf II. ROTATIONAL MOTIONS}

\medskip

Let the closed string with three material points uniformly rotate
(preserving its form in time) in a plane $xy$ around the origin of
coordinates. The quark trajectories in $(2+1)$-dimensional Minkowski
space are the screw lines. For this motion one can choose on the
world surface a parametrization with screw lines $\sigma={}$const and
with an uniform growth of $\tau$ along these lines. In these
coordinates on the quark trajectories
\begin{eqnarray}
\sigma_i(\tau)=\,\mbox{const},\quad i=0,1,2,3,\qquad\nonumber\\
|V_i|=\sqrt{\dot X^2}\big|_{\sigma=\sigma_i}=C_i=\,\mbox{const},
\quad i=1,2,3.
\end{eqnarray}

All four functions $\sigma_i(\tau)$ are fixed simultaneously,
$\sigma_1$
and $\sigma_2$ --- in the form (7) non-limiting a generality.

Let coordinate curves $\tau={}$const be orthogonal trajec\-tories
to the specified lines $\sigma={}$const and conditions (5)
be satisfied. These lines $\tau={}$const (don't coinciding with
sections $t={}$const) are not closed, but a connection between
$\tau$ and $\tau^*$ in the closure condition (2) is very sim\-ple:
$\tau^*=\tau+{}$const.
It is a consequence of the symmetry of this motion ---
the world surface in Minkowski space coincide with itself after
a rotation about $t=x^0$\,-\,axis with simultaneous translation along
this axis.

Under these circumstances and conditions (5), (9) the third
Eq.~(4) takes the form
\begin{eqnarray}
m_3C_3^{-1}\ddot X^\mu(\tau,\sigma_0)-\gamma
X'{}^\mu(\tau,\sigma_0+0)\;\;\nonumber\\
{}+\gamma X'{}^\mu(\tau^*,\sigma_3-0)=0,\quad
\tau^*=\tau+\mbox{const}.
\end{eqnarray}

A solution of the string oscillatory equation (6) sati\-sfying the
conditions (5), (7)\,--\,(9) may be found by the Fourier method:
$X^\mu=\sum_k e_k^\mu u_k(\sigma)\,T_k(\tau)$.
The functions $u_k(\sigma)$ and $T_k(\tau)$ with the same $k$
as a consequence of Eq.~(6) are linear functions or harmonics
with the same frequency $\omega$.
Taking into account the described above pro\-perties of the
rotational motion and its parametrization one
can find the Fourier series for $X^\mu$ in $(2+1)$ Minkowski space
(with the unique frequency $\omega$) in the form
\begin{eqnarray}
X^\mu=\bigl\{t_0+a\tau+b\sigma;\,
u(\sigma)\cos\omega\tau-\tilde u(\sigma)\sin\omega\tau;\nonumber\\
u(\sigma)\sin\omega\tau+\tilde u(\sigma)
\cos\omega\tau\bigr\}.
\end{eqnarray}

The functions $u(\sigma)$ and $\tilde u(\sigma)$ are continuous
in $[\sigma_0,\sigma_3]$, may have discontinuities of
derivatives at $\sigma=0$, $\sigma=\pi$ and in the segments
$[\sigma_{i-1},\sigma_i]$ are:
\begin{eqnarray}
u(\sigma)=\left\{\begin{array}{ll}
A_0\cos\omega\sigma+B_0\sin\omega\sigma,&\sigma\in[\sigma_0,0],\\
A_1\cos\omega\sigma+B_1\sin\omega\sigma,&\sigma\in[0,\pi],\\
A_2\cos\omega\sigma+B_2\sin\omega\sigma,&\sigma\in[\pi,\sigma_3];
\;\;\end{array}\right.\\
\quad\tilde u(\sigma)=\tilde A_i\cos\omega\sigma+
\tilde B_i\sin\omega\sigma,\;\;\sigma\in[\sigma_i,\sigma_{i+1}].
\nonumber\end{eqnarray}

Let the functions $e^\mu u(\sigma)\,T(\tau)$ and
$e^\mu\tilde u(\sigma)\,T(\tau)$ (with $T=\cos\omega\tau$ or
$T=\sin\omega\tau$) satisfy the boundary con\-di\-tions (8)
independently. With the continuity condi\-tions at $\sigma=0$
and $\sigma=\pi$ it results in 4 equations both for $u$ and
$\tilde u$ which may be presented in the form solved with respect
to $A_1\equiv A$ and $B_1\equiv B$ (the same formulae express
$\tilde A_i,\tilde B_i$ by $\tilde A_1\equiv\tilde A$ and
$\tilde B_1\equiv\tilde B$):
\begin{eqnarray}
A_0=A,\qquad B_0=h_1A+B,\nonumber\\
A_2=(1+h_2c_1s_1)\,A+h_2s_1^2B,\,\\
B_2=-h_2c_1^2A+(1-h_2c_1s_1)\,B.\nonumber
\end{eqnarray}
Here $\quad c_1=\cos\pi\omega,\;\;s_1=\sin\pi\omega,
\quad h_i=\omega m_i/(\gamma C_i)$.

Under relations (13) solution (11)\,--\,(12) satisfies
conditions (8). Substitution Eqs.~(11)\,--\,(12) into the second
of the orthonormality conditions (5) results in three equations
\begin{equation}
A_i\tilde B_i-\tilde A_iB_i=ab/\omega^2,\quad i=0,1,2.
\end{equation}
But among Eqs.~(14) only one is independent, for example, with
$i=1$. If it's satisfied and the relations (13) take place ---
two other conditions (14) are satisfied too. And v.v. substitution
(11)\,--\,(12) into the first condition (5) results in
$A_i^2+B_i^2+\tilde A_i^2+\tilde B_i^2=(a^2+b^2)/\omega^2$
--- three independent equations.
Transform this system with taking into account Eqs.~(13) in
the following equivalent form:
\begin{eqnarray}
&A^2+B^2+\tilde A^2+\tilde B^2=(a^2+b^2)/\omega^2,&\\
&h_1(A^2+\tilde A^2)+2(AB+\tilde A\tilde B)=0,&\\
&\lambda_1(A^2+\tilde A^2)=\lambda_2(B^2+\tilde B^2).&
\end{eqnarray}
Here $\lambda_1=(h_1h_2-2)\,c_1s_1+h_1(1-2c_1)-h_2c_1^2$
and\linebreak
$\lambda_2=h_2s_1^2-2c_1s_1$.

Expression (11) is a solution of the given problem if the last
necessary conditions (2) and (10) are satisfied. Denote
$-\theta/\omega$ the constant in Eq.~(10):
\begin{equation}
\tau^*=\tau-\theta/\omega,\quad\theta=(\sigma_3-\sigma_0)\,
\omega b/a=D\omega b/a.
\end{equation}

The expression for $\theta$ results from the substitution of\linebreak
$X^0=t_0+a\tau+b\sigma$ into the closure condition (2).
The angle $\theta$ has the following geometrical sense:
$\theta$ is the phase shift on a screw trajectory of the third quark
between the beginning (at $\sigma=\sigma_0$) and the end (at
$\sigma=\sigma_3$) of an unclosed coordinate line $\tau={}$const.

Substitute Eqs.~(11)\,--\,(13), (18) into the closure (2) and
boundary (10) conditions with $\mu=1,2$. Values of $u$, $\tilde u$
and their derivatives at $\sigma=\sigma_0$ and $\sigma_3$
express through $A$, $B$, $\tilde A$, $\tilde B$ by Eqs.~(13),
for example:
$u(\sigma_3)=\big[\cos\omega\sigma_3-h_2c\sin\omega(\sigma_3-\pi)\big]\,A+
\big[\sin\omega\sigma_3-h_2s\sin\omega(\sigma_3-\pi)\big]\,B$.

Equating the similar terms with $\cos\omega\tau$ and
$\sin\omega\tau$ in 4 Eqs.~(2), (10) with $\mu=1,2$ we obtain 8
homogeneous equations with respect to $A$, $B$, $\tilde A$,
$\tilde B$ which reduce to 4 pairs of coinciding ones.
For the sake of simplicity and explication of its intrinsic
structure we write this homo\-geneous system with the martix notation:
\begin{equation}
M_1\alpha=M_2\beta,\qquad M_3\alpha=M_4\beta.
\end{equation}
Here
$\;\alpha=\left(\begin{array}{c}A\\ \tilde A\end{array}\right)$,
$\,\beta=\left(\begin{array}{c}B\\ \tilde B\end{array}\right)$
and matrices
\begin{eqnarray*}
M_1\!&=&\!(h_1s_0-c_0)\,I+(c-h_2c_1s_2)\,U,\\
M_2\!&=&\!-s_0I-(s-h_2s_1s_2)\,U,\\
M_3\!&=&\!\bigl[(1-h_1h_3)\,s_0+(h_1+h_3)\,c_0\bigr]I+(s+h_2c_1c_2)
\,U,\\
M_4\!&=&\!(h_3s_0-c_0)\,I+(c-h_2s_1c_2)\,U
\end{eqnarray*}
are linear combinations of the identity matrix $I$ and the matrix
$U=U(\theta)=\left(\begin{array}{cc}\cos\theta
&\sin\theta\\ -\sin\theta &\cos\theta\end{array}\right)$.

The coefficients are
\begin{eqnarray*}
&c_i=\cos\omega d_i,\;\;s_i=\sin\omega d_i;\;\;
c=\cos\omega\sigma_3,\;\;s=\sin\omega\sigma_3;&\\
&d_i=\sigma_{i+1}-\sigma_i:\quad d_0=-\sigma_0,\;\;d_1=\pi,&\\
&d_2=\sigma_3-\pi;\quad D=\sigma_3-\sigma_0=d_0+d_1+d_2.&
\end{eqnarray*}

Taking into account mutual commutability of $M_k$ one can
exclude $\alpha$ or $\beta$ from the system (19)
\begin{equation}
M\alpha=0,\qquad M\beta=0,
\end{equation}
$M=M_1M_4-M_2M_3=I+U^2-FU=(2\cos\theta-F)\,U$
(an equality $I+U^2(\theta)=2\cos\theta\cdot U$ is used).
The parameter $F$ may be transformed to the simple form
\begin{center}
$F=2\cos\omega D-\sum\limits_ih_i\sin\omega D
+\sum\limits_{i<j}h_ih_j s_i\sin\omega(d_{i-1}+d_j)$\\
${}-h_1h_2h_3s_1s_2s_0=G_1+G_2+G_3-G_1G_2G_3$
\end{center}
through the following notation:
\begin{equation}
G_i=\frac{h_is_{i-1}s_i-\sin\omega(d_{i-1}+d_i)}{s_{i+1}}.
\end{equation}

The notation here is cyclically equivalent:
$d_{i+3}\equiv d_i$, $s_{i+3}\equiv s_i$, $G_{i+3}\equiv G_i$,
for example, $d_3\equiv d_0$, $s_4\equiv s_1$.

Homogeneous systems (20) have a desirable non-trivial solution
if and only if $\det M=(2\cos\theta-F)^2=0$, i.e.,
\begin{equation}
2\cos\theta=G_1+G_2+G_3-G_1G_2G_3.
\end{equation}

Under condition (22) the matrix $M=0$ and an arbitrary
non-zero column $\alpha$ or $\beta$ is its eigenvector. It is
connected with the rotational symmetry of the problem. So one can
choose an optional pair $A$\,\&\,$\tilde A$, $B$\,\&\,$\tilde B$
or $A$\,\&\,$B$ and determine two other constants from Eq.~(19)
(under condition (22) two systems (19) are equivalent), in
particular:
\begin{equation}
\tilde A=-K\,(h_1A+2B),\quad\tilde B=K\,(2HA+h_1B),
\end{equation}
where
\vspace{-3ex}
\begin{equation}
K=\frac{s_0s_1(G_2G_3-1)}{2s_2\sin\theta},\quad
H=\frac{1+h_1^2K^2}{4K^2}.
\end{equation}

Values (23) must obey conditions (14)\,--\,(17) descending
from the orthonormality conditions (5). Substitution of (23) in
Eqs.~(16) and (17) after transformations results in relations
\begin{equation}
\frac{G_{i+1}-G_i}{G_iG_{i+1}-1}=\frac{\sin\omega(d_{i-1}
-d_{i+1})}{s_i},\quad i=1,2,3.
\end{equation}
One of these equations ($i=2$) is a consequence of (16),
the second --- of (17) and the third --- of the previous two.

Substitution of (23) in Eqs.~(14) and (15) after transfor\-mations
with taking into account (18), (21)\,--\,(25) results in two
equations which may be written in the form
\begin{eqnarray}
&a^2=2KD\omega^3\theta^{-1}(HA^2+h_1AB+B^2),&\\
&\displaystyle\frac{D\omega\theta}{D^2\omega^2+\theta^2}=
\frac{2K}{1+(4+h_1^2)\,K^2}&
\end{eqnarray}
with $K$ from (24).

The latter equation determines a set of acceptable frequencies
$\omega$ if the parameters $G_i$, $d_i$ and $\theta$ are given.
All these parameters defining a rotational motion of the model
(except for translations and a scale factor) are related by the
system of non-linear equations (21), (22), (24), (25) and (27).

The simplest way to construct solutions of the consi\-dered problem
is to start with fixing three parame\-ters $G_1$, $G_2$, $G_3$ as
initial data. On the next step we deter\-mine the angle $\theta$ by
Eq.~(22). The result of this procedure isn't unique --- for
an every triplet $G_i$ one can find a countable set of values
$\theta=\theta_{j_1}$. Further, the lengths $d_i$
are defined from Eqs.~(25) by the following two steps (the
value $d_1=\pi$ was already chosen):
\begin{eqnarray*}
&\displaystyle\delta=d_0-d_2=\frac1\omega\Bigl[(-1)^{j_2}\arcsin s
\frac{G_2-G_1}{G_1G_2-1}+\pi j_2\Bigr],&\\
&\displaystyle d_0=\frac1\omega\Bigl[\arctan\frac{\sin\omega(\delta+
\pi)}{\cos\omega(\delta+\pi)-\frac{G_1-G_3}{G_1G_3-1}}+\pi j_3\Bigr]&
\end{eqnarray*}
and $d_2=d_0-\delta$ with arbitrary integer $j_2$, $j_3$.
Substi\-tu\-tion of $G_i$, $d_i$, $\theta$ and $K$
into Eq.~(27) results in a countable set of frequencies
$\omega$. The latter equation is solved numerically by the secant
method [14].

After a choice of the amplitudes $A$ and $B$
one can determine the values $\tilde A$, $\tilde B$, $a$, $b$
correspondingly by Eqs.~(23), (26), (18) and through
Eqs.~(12)\,--\,(13) --- the world surface (11).

To investigate the constructed world surface one can consider its
section $t=t_0={}$const --- a ``photograph" of the string position
at time moment $t_0$. These sections (curvilinear triangles) are
shown in Fig.~2.  Without
limiting generality $A=0$ is supposed in these examples --- a
transition to another ``gauge" with $A\ne0$ doesn't change the form of
such a curve, but only rotates it.

A parametrization of these curves is
\begin{eqnarray}
&x=u(\sigma)\cdot\cos\frac\theta D\sigma+
\tilde u(\sigma)\cdot\sin\frac\theta D\sigma,&\\
&y=-u(\sigma)\cdot\sin\frac\theta D\sigma+
\tilde u(\sigma)\cdot\cos\frac\theta{D}\sigma,&\nonumber
\end{eqnarray}
in particular, for two sides of the ``triangle" ($A=0$)
$$u=B\sin\omega\sigma,\;\,\tilde u=BK(h_1\sin\omega|\sigma|-
2\cos\omega\sigma),\;\sigma\in[\sigma_0,\pi].$$

The curve (28) is composed of three segments of a hypocycloid
joined at non-zero angles in three points (the quark positions).
Hypocycloid is the curve drawing by a point of a circle (with radius
$r$) that is rolling in another fixed circle with larger radius
$R$ [15]. In the case (28) a relation of these radii
\begin{equation}
R/r=2/(1-|b|/a)=2\big/\big(1-|\theta\big/(D\omega)|\big)
\end{equation}
is irrational in general.

Differentiating Eqs.~(28) results in the following fact: the curve
(28) (for its smooth segments) is the hypocycloid if and only if
the parameters of the curve are bounded by Eq.~(27).

The curves in Fig.~2 rotate in the $xy$ plane at the
angular velocity $\Omega=\omega/a$ where $a$ is determined by Eq.~(26);
three quarks move at speeds
\begin{equation}
v_i=\sqrt{\frac{\theta s_{i-1}s_i(G_{i-1}G_{i+1}-1)}
{\omega Ds_{i+1}\sin\theta}},\quad i=1,2,3,\end{equation}
along circles with radii $R_i=v_i/\Omega=av_i/\omega$.

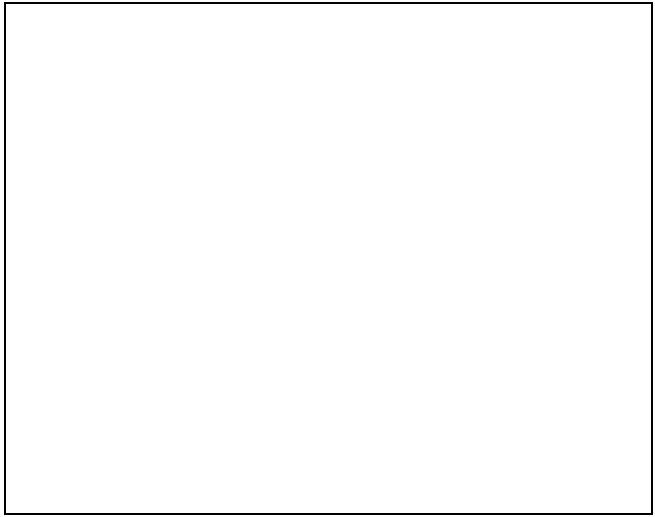
\begin{figure}[t]
\begin{picture}(86,68)
\put(0,0){\line(1,0){86}} \put(0,68){\line(1,0){86}}
\put(0,0){\line(0,1){68}} \put(86,0){\line(0,1){68}}
\end{picture}
\caption{The simple states with various rotational rates
for the system with $m_1=4$, $m_2=2$, $m_3=3$.}
\end{figure}

A free choice of the mentioned integer parameters $j_1$, $j_2$,
$j_3$, $l$ results in a very large number of different motions of
the system distinguishing from each other by their topological
structure. A motion or state of the system we'll denominate  {\sl
``simple"} if the position of the string (section $t={}$const) is
a curvilinear triangle with smooth sides (Fig.~2).
In the opposite case if there are come singular massless points
on the sides of the ``triangle" we'll denominate the state
{\sl``exotic"}. These singular points move at the
speed of light.

The motion of the system is simple if its parameters satisfy
the following conditions:
\begin{eqnarray}
|\theta|<\pi,\quad|\omega|d_i<\pi,\quad G_i>1,\quad i=1,2,3,\nonumber\\
G_1+G_2+G_3-G_1G_2G_3>-2.\end{eqnarray}
In particular, if two quark masses are equal, for example, $m_2=m_3$,
the conditions (31) for $G_i$ take the form $G_1>1$,
$1<G_2=G_3<1+2/G_1$. And in a symmetric case $m_1=m_2=m_3$
the limitation (31) for the simple states is $1<G_1=G_2=G_3<2$.

The dependence of a form of the curvilinear triangle on its
rotational speed is shown in Fig.~2 for the case of the
simple motion of the system with fixed quark masses $m_i$. In this
figure the ``photographs" of the same system in various
rotational states are placed, the higher quark speeds --- the larger
size of the curvilinear triangle. The dependence $R_i(v_i)$ for
$m_i/\gamma={}$const is too sharp so the smallest (inner) triangle
in Fig.~2 is 5 times magnified, and the largest is 7 times  diminished
in comparison with
the natural size (natural --- is in the case $m_i/\gamma={}$const).
For the middle curve $\gamma=1$ is taken.

The speed of rotation could be measured by anyone of the parameters
$v_i$, $\omega$, $\Omega$, $\theta$, $G_i$, the energy $E$, the
angular momentum $J$ (Sec.~3), etc.
Some rounded values of these and others parameters (in particular,
the minimal $v_1$ and maximal $v_2$ quark speeds in the case
of different $m_i$ for the simple motions in Fig.~2 are presented
in the following table:
\smallskip

\begin{center}
\begin{tabular}{|c|c|c|c|}  \hline
&{\scriptsize inner}&{\scriptsize middle}&{\scriptsize outer}\\ \hline
$G_1$&  1.15& 1.7& 2.2\\
$G_2$&  1.041& 1.312& 1.69\\
$\omega$& 0.143& 0.386& 0.752\\
$\theta$& 0.139& 0.923& 2.328\\
$d_2$& 3.955& 3.493& 3.222\\
$v_1$& 0.145& 0.497& 0.909\\
$v_2$& 0.354& 0.711& 0.954\\
$E$&  9.832& 16.38& 95.61\\
$J$&  0.231& 7.005& 520.0\\
\hline
\end{tabular}
\end{center}
\smallskip

In the symmetric case that is considered in Ref.~[9] for the simple states
the following parameters are equal: $G_1=G_2=G_3=G$, $d_1=d_2=d_3=\pi$,
$v_1=v_2=v_3=v$, and a value of the parameter
$G$ in the interval $(1,2)$ is the measure of rotational rate.

In the case with different masses $m_i$ given as initial data
(Fig.~2) the mentioned procedure needs some complement.
The given quark masses are connected with the others parameters
of the system by the expressions
\begin{equation}
m_i=\frac\gamma\omega C_ih_i=
\frac\gamma\omega h_ia\sqrt{1-v_i^2},\end{equation}

The values of the parameters for the rotational states in
Fig.~2 were calculated as follows: a value of $G_1$ was
chosen as a measure of rotational rate, $G_2$ and $G_3$ were taken
as tentative at the first step of an iteration. After realization
of the mentioned procedure of determination of $\theta$, $d_i$,
$\omega$, etc., the masses (32) (or relations $m_i/m_1=
(h_i/h_1)\sqrt{(1-v_i^2)/(1-v_1^2)}$, $i=2,3$) were found and
compared with the given values. The two-dimensional secant method
[14] was applied in this iterative process.

The simple states in Fig.~2 demonstrate the fol\-lowing
asymptotics in a non-relativistic and an ultra\-relativistic limits.
If the quark velocities $v_i$, the system energy $E$, the momentum $J$
and the values $\omega$ and $\theta$ decrease, the curvilinear
triangle tends to a rectilinear triangle. A form of the latter depends
on an answer the question: is the triangle inequality for the quark
masses $m_1$, $m_2$, $m_3$ satisfied?

If this inequality is satisfied, i.e., each of the quark masses $m_i$
is less then a sum of two others (Fig.~2)
in the non-relativistic limiting case the parameters $\omega$,
$\theta$, $v_i$, $R_i$ tend to 0, $G_i\to1+0$ for all $i=1,2,3$;
the triangle tends to the rectilinear one, and lengths of its
sides $l_{ij}$ (between $i$-th and $j$-th quark) in this limit
are proportional to associated $d_i$ and opposite quark masses:
\begin{equation}
\frac{l_{12}}{d_1}=\frac{l_{23}}{d_2}=\frac{l_{31}}{d_0},
\quad\frac{d_1}{m_3}=\frac{d_2}{m_1}=\frac{d_0}{m_2},
\qquad  v_i\to0.\end{equation}

If one of these masses is larger then a sum of two others, for
example $m_1>m_2+m_3$, in the low energy limit $\theta\to0$
the obtuse angle at the corner with the largest mass $m_1$ tends to
$\pi$ and the triangle tends to a rectilinear segment. This limit
is attained with $\theta=0$, $R_1=0$, $v_1=0$, $G_2=G_3=1+0$ and
non-zero values of $\omega$, $v_2$, $v_3$, $G_1>1$. These
limiting values are connected by the equations resulting from
Eqs.~(21), (24)\,--\,(27), (30), (32):
\begin{eqnarray}
&v_2=s_1,\quad d_2=d_0+d_1,\quad G_1=1+2d_2\omega s_0s_1s_2^{-1},&
\nonumber\\
&v_3=s_0,\quad\displaystyle \frac{m_2}{m_3}=\frac{c_1^2s_0}{c_0^2s_1},
\quad\frac{m_1}{m_3}=d_2\omega s_0s_1+s_2.&\end{eqnarray}

If $v_2$ and $v_3$ become less then the limiting values (34), the
heaviest quark occupies a position at the rotational centre and the
string rotates as the rectilinear segment. It looks like the string
mesonic model with two light quarks, bounded by two relativistic
strings (details in Sec.~3) with a supplement --- the heavy quark
at rest.

In the ultrarelativistic limit $v_i\to1$ for the simple states the
values $d_0$ and $d_2$ tend to $d_1=\pi$, $|\omega|\to1-0$,
$|\theta|\to\pi-0$, and the curvilinear triangle tends to a
hypocycloid with three arcs (deltoid)
\begin{equation}
\array{l} x=B\big(2\sin\frac23\sigma-\sin\frac43\sigma\big),\\
y=-B\big(2\cos\frac23\sigma+\cos\frac43\sigma\big),\\
\endarray\quad\sigma\in[-\pi,2\pi].\end{equation}

A form of the limiting curve (35) doesn't depend on the (fixed)
values $m_1$, $m_2$, $m_3$.
So one can deduce Eqs. (35) by the simplest way in the
symmetric case $m_1=m_2=m_3$. In this case the ultrarelativistic
limit $v_i\to1$ corresponds to a limit $G_i=G\to2-0$. Substitution
of expres\-sions $\omega=1-\delta$, $G=2-g^2$ with infinitesimals
$\delta$, $g$ into Eqs.~(24), (27) results in the limiting relation
$\frac3{10}=\lim\limits_{g\to0}\pi\delta
g\,(g^2+\pi^2\delta^2)^{-1}$.
The root
$\lim\limits_{g\to0}\pi\delta/g=3$ of this
square equation corresponds to the desirable physical case $m_i>0$.
The following terms of expansion $\omega$ and $\theta$ in (27) are:
\begin{equation}
\omega\simeq1-\frac3\pi g+\frac{15}{8\pi}g^3,
\quad\theta\simeq\pi-3g,\;\;g\to+0.\end{equation}
Substitution of (36) in Eqs.~(22) and (13) results in limiting
expressions at $g\to+0$ (in the case $A=0$)
$\,\,u(\sigma)=B\sin\sigma,\quad\tilde u(\sigma)=
-3B\cos\sigma,\quad\sigma\in[-\pi,2\pi]$ and the world surface (11)
\begin{eqnarray}
X^\mu=B\bigl\{t_0+3\tau+\sigma;\,\sin\sigma\cos\tau+3\cos\sigma
\sin\tau;\nonumber\\
\sin\sigma\sin\tau-3\cos\sigma\cos\tau\bigr\}.\end{eqnarray}
A section of this surface $t={}$const is the hypocycloid (35).

Let's consider a situation, where the condition of ``simplicity" (31)
are not satisfied. Such a motion was denominated as exotic. Its
world surface has peculiarities $\dot X^2=X^{\prime2}=0$ on the
world lines of singular points (cusps) of the hypocycloid (28)
which move at the speed of light.

There are many types of exotic motion differing from each other
by the number and positions of these peculiarities.
These topological configurations of the exotic states may be
classified by investigation of the massless $m_i\to0$ or
ultrarelativistic $v_i\to1$ limit. In this limit for the exotic states
Eqs.~(22)\,--\,(27), (30) result to expressions
$$\lim\limits_{m_i\to0}\frac{|\omega| d_i}\pi=1+n_i,\quad
\lim\limits_{m_i\to0}h_i=0,\quad\lim\limits_{m_i\to0}2K=\frac nk,$$
where
\begin{equation}
n=\lim\limits_{m_i\to0}\frac{|\omega|D}\pi=n_1+n_2+n_3+3,\quad
k=\lim_{m_i\to0}\frac\theta\pi.\end{equation}
Here $n_1$ is the number of singular points between the 1-st and the
2-nd quark, $n_2$ --- between 2 and 3, $n_3$ --- between 3 and 1,
$k$ is an integer.

Substitution of these expressions into Eq.~(23) with $A=0$ results in
the following limiting form of the world surface for all parts of the string
as a generalization of Eq.~(37):
\begin{eqnarray}
X^\mu=B\bigl\{n\tau+k\sigma;\,k\sin\sigma\cos\tau+
n\cos\sigma\sin\tau;\nonumber\\
k\sin\sigma\sin\tau-n\cos\sigma\cos\tau\bigr\}.\end{eqnarray}
Here $\sigma\in[0,\pi n]$, the integer parameters (38) $n$ and $k$
are restricted by the conditions
\begin{equation}
n\geq2,\quad|k|\leq n-2,\quad n-k\mbox\quad\mbox{is even}.
\end{equation}
For the simple states (31) $n=3$, $|k|=1$.

Note that world surfaces (39) describe motions of a closed
massless relativistic string. Expression (39) is a solution of
Eq.~(6), satisfies the orthonormality conditions (5) and the closure
condition $\,X^\mu(\tau,0)=X^\mu(\tau-\pi k,\pi n)$
with $n\geq2$ and $k$ restricted by (40).

A section $t={}$const of world surface (39) is a closed
hypocycloid with rational relation of the two radii
$$R/r=2n\big/(n-|k|)$$
(compare with Eq.~(29)). If $|k|=n-2$, this relation equals $n$
and the curve has no selfintersections. If $|k|\leq n-4$, the
hypocycloid is starlike. The singular points of these hypocycloids
move at the speed of light.

Topological types of rotational motions of the considered system
may be exhaustively classified by pointing out a set of the mentioned
integer parameters $(n,k;n_1,n_2,n_3)$ which are connected by
Eq.~(38) and satisfy the inequalities (40).

The states of the system differing from each other only by changing
$k$ to $-k$ should be interpreted as the same topological type. It
results from the fact that replacement $\theta$ to $-\theta$ in
Eqs.~(18)\,--\,(28) changes only the bypass direction of the
curvilinear ``triangle".

In the case $k=0$ (it's possible for even $n$) the exotic state has
a form of uniformly rotating rectilinear string that is folded
some times.
The simplest of these states $n=2$, $k=0$ is the case of coincidence
of two quarks (one of $d_i$ equals 0). In this state the model
``triangle" practically reduces to the quark-diquark one with the
quark and diquark, connected by the double string with the tension
$2\gamma$. This rectilinear segment is the particular case of the
hypocycloid with $R/r=2$.

If $n\ge4$, $k=0$ the quarks and the massless peculiarities
$\dot X^2=0$ are situated at the fold points. In this case in Eq.~(11)
$b=\theta=0$, $\tilde u(\sigma)=u(\sigma)\cdot{}$const. These
states have analogs in the meson string model with massive ends.
A solution [16]
\begin{equation}
X^\mu=\bigl\{\alpha\tau;\,Bu_n(\sigma)\cos\omega_n\tau;\,
Bu_n(\sigma)\cos\omega_n\tau\bigr\}
\end{equation}
describes a rotation of $n-1$ times folded rectilinear open string.
Here $u_n(\sigma)=\cos\omega_n\sigma-\omega_nQ_1^{-1}
\sin\omega_n\sigma$, $\sigma\in[0,\pi]$,
$Q_i=\gamma m_i^{-1}\sqrt{\dot X^2}\bigg|_{\sigma=\sigma_i}\!=$\,const
and $\omega_n$ is $n$-th positive root of the transcendental equation
$\tan \pi\omega=(Q_1+Q_2)\,\omega\big/(\omega^2-Q_1Q_2)$.

\bigskip

\centerline{\bf III. ENERGY AND ANGULAR MOMENTUM}
\centerline{\bf OF ROTATIONAL STATES}

\medskip

In this section possibility of application of considered solutions
for description of baryon states on the Regge trajectories is
briefly discussed. The Regge trajectory includes states of baryons
with the same quark composition and almost the same set of quantum
numbers. This trajectory is linear dependence (without satisfactory
theoretical explanation) between square of mass or rest energy of
the particle $M^2=E^2$ and its spin or angular momentum $J$:
$J=\alpha'E^2+\alpha_0$.

Let's find a connection between the energy $E$ and angular momentum $J$
of the rotational state (11) of the baryonic model ``triangle" on the
classic level. The same problem for the string model of meson is
solved in Ref.~[3,8].

In accordance with [2,3] consider new parameters $t,\sigma$ on the
world surface, where $t=X^0$ is time, $\sigma$ is the former
parameter. The Lagrangian in action (1) is
$\Lambda=-\gamma\int_{\sigma_0}^{\sigma_3}L(\bar X_t,\bar X_\sigma)
,d\sigma-\sum_{i=1}^3m_i\sqrt{1-\bar X_t^2(t,\sigma_i)}$, where
$L=\big[(\bar X_t\bar X_\sigma)^2+\bar X_\sigma^2(1-\bar X_t^2)
\big]^{1/2}$,
$\bar X_t=\partial_t\bar X$, $\bar X_\sigma=\partial_\sigma\bar X$,
$\bar X=\big\{X(t,\sigma),\,Y(t,\sigma)\big\}$ --- 2D-vector,
the scalar product is Euclidean.

In the co-ordinates $t,\sigma$ the orthonormality conditions (5)
aren't satisfied, so the canonical momentum
$\bar P(t,\sigma)=\delta\Lambda/\delta\bar X_t=-\gamma\big[(\bar X_t
\bar X_\sigma)\,\bar X_\sigma-\bar X_\sigma^2\bar X_t\big]\big/L+
\sum_{i=1}^3m_i\bar X_t(1-\bar X_t^2)^{-1/2}\delta(\sigma-\sigma_i)$
is non-linear with respect to $\bar X_t$.

The energy of the system
$E=\int_{\sigma_0}^{\sigma_3}(\bar X_t\bar P)\,d\sigma-\Lambda={}$
$\gamma\int_{\sigma_0}^{\sigma_3}L^{-1}\bar X_\sigma^2\,d\sigma+
\sum_{i=1}^3m_i\big/\sqrt{1-v_i^2}$
has the form
\begin{equation}
E=\gamma D\frac{a^2-b^2}a+\sum_{i=1}^3\frac{m_i}{\sqrt{1-v_i^2}},
\end{equation}
where $v_i^2=\bar X_t^2(t,\sigma_i)$. The following expressions
resulting from Eqs.~(11)\,--\,(17) were used in the calculations:
$\bar X_\sigma^2=(a^2-b^2)\,(1-\bar X_t^2)=
(a^2-b^2)\,b^{-1}(\bar X_t\bar X_\sigma)=(a^2-b^2)\,a^{-1}L$.
The parameters $D$, $a$, $b=a\theta/(D\omega)$ and $v_i$ are defined
by Eqs.~(18), (26) and (30).

The angular momentum
$J=\int_{\sigma_0}^{\sigma_3}(XP_y-YP_x)\,d\sigma$
of the state (11) is calculated by the similar way
\begin{equation}
J=\frac a{2\omega}\bigg(E-\sum_{i=1}^3m_i\sqrt{1-v_i^2}\bigg).
\end{equation}
The latter relation between $E$, $J$ and the angular frequency
$\Omega=\omega/a$ has almost the same form as in the string model
of meson [3].

Expressions (42) and (43) set an implicit non-linear connection
between $E$ and $J$ of the considered system. A form of this
connection depends on the topological type $(n,k;n_1,n_2,n_3)$
of the state of the system.

In Fig.\,3 the results of numerical calculation the dependence
$J$ on $E^2$ are shown for various states of the systems with fixed
$m_2=m_3=0.3$, $\gamma=3/(16\pi)$. Such a choice of $\gamma$
approximately corresponds to the experimental value
$\alpha'\simeq1$\,GeV${}^{-2}$. One can suppose conditionally that
$E^2$ in Fig.\,6 is measured in GeV${}^2$ and $J$ --- in units
$\hbar$.

Curves 1,\,2,\,3 and 4 describe the simple motions of the system
correspondingly with $m_1=0.05$, $m_1=0.3$, $m_1=0.6$ and $m_1=1$;
curve 5 --- the exotic state of the symmetric system with equal
masses $m_1=m_2=m_3=0.3$ and with topological type of this state
$(6,2;1,1,1)$.

The symbol ``T" on curve 4 shows the point of transformation of
the triangular configuration of this system with $m_1=1>m_2+m_3=0.6$
to the rectilinear configuration. This point of transformation
corresponds to satisfying of Eqs.~(34).

The analysis shows that
in the non-relativistic limit the asymptotic behaviour
of the function $J(E)$ depends on satisfaction of the triangle
inequality between three masses $m_i$. If each mass is less than
the sum of two others, the limiting relations (33) for the simple
state take place, and energy (42) of this state has the form
$E=\sum_{i=1}^3m_i+\frac32m_1m_2m_3^{-1}\pi^2\omega^2+o(\omega^2)$,
where $\omega$ is an infinitesimal. The considered
asymptotic relation in this case is
$$J\simeq\left(\frac23\right)^{\!3/2}\!\frac{\sqrt{m_1m_2m_3}}
{\gamma\,(m_1+m_2+m_3)}\big(E-{\textstyle\sum\limits_{i=1}^3m_i}
\big)^{3/2},\quad v_i\ll1.$$

\begin{figure}[tb]
\begin{picture}(86,68)
\put(0,0){\line(1,0){86}} \put(0,68){\line(1,0){86}}
\put(0,0){\line(0,1){68}} \put(86,0){\line(0,1){68}}
\end{picture}
\caption{Dependence $J(E^2)$ for the simple motions with various
$m_i$ (1\,--\,4) and for the exotic state (5).}
\end{figure}
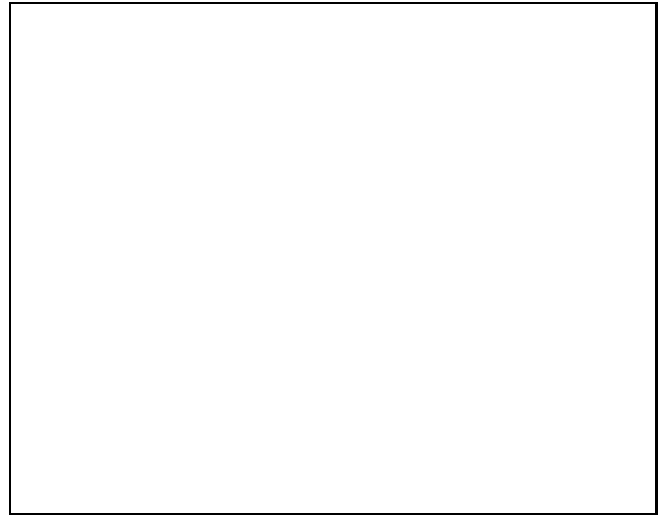

If one of the masses, for example, $m_1$ is larger then the sum
of two others (curve 4 in Fig.~3), then the
non-relativistic asymptotic case describes a rotation of rectilinear
double string with two masses $m_2$ and $m_3$ at the ends and
the mass $m_1$ at the rotational center. In this limit for the
simple state the following expression take place
$$J\simeq\left(\frac23\right)^{\!3/2}\!\frac1{2\gamma}\sqrt{
\frac{m_2m_3}{m_2+m_3}}\big(E-{\textstyle\sum\limits_{i=1}^3m_i}
\big)^{3/2},\quad v_i\ll1.$$
It looks like the formula in Ref. [3] for the string model of meson
and may be deduced from solution (41), but the tension
of the string equals $2\gamma$.

The exponent $3/2$ is the same for both cases considered. So
graphs 1\,--\,4 in Fig.~3 have similar forms and curve 4 in
the vicinity of the transformation point is rather smooth.

In the opposite ultrarelativistic limit $v_i\to1$, $E\to\infty$,
$J\to\infty$ the analysis of dependence $J(E)$ includes
substituting limiting formulae (38) and expressions
$\omega d_i=\pi(n_i+1)-\delta_i$,
$\theta=\pi k(1-\delta_\theta)$, $\sqrt{1-v_i^2}=\varepsilon_i$
with the infinitesimals $\delta_i$, $\delta_\theta$,
$\varepsilon_i$ (generalization of Eq.~(36)) into
Eqs.~(22)\,--\,(30), (42) and (43). Expansion in series in
Eqs.~(25), (27) and (30) results in the following relations
between the infinitesimals:

\begin{eqnarray*}
&\displaystyle
h_i\simeq2\frac{\sqrt{n^2-k^2}}n\varepsilon_i\Big(1+
\frac{n^2-2k^2}{2(n^2-k^2)}\varepsilon_i^2\Big),&\\
&\displaystyle\frac{\delta_i}{\sqrt{m_i}+\sqrt{m_{i+1}}}\simeq
\frac{\delta_j}{\sqrt{m_j}+\sqrt{m_{j+1}}}\simeq
\frac n{\sqrt{n^2-k^2}}\frac{\varepsilon_i}{\sqrt{m_i}},&\\
&\displaystyle\sum\limits_{i=1}^3\delta_i\simeq\frac{2nm_1^{-1/2}}
{\sqrt{n^2-k^2}}\big({\textstyle\sum\limits_{i=1}^3m_i^{1/2}}\big)
\,\varepsilon_1+\frac{nm_1^{-3/2}}{(n^2-k^2)^{3/2}}\times&\\
&\displaystyle\left(\frac{n^2-2k^2}2
m_1{\textstyle\sum\limits_{i=1}^3m_i^{1/2}}-\frac{n^2-6k^2}6
{\textstyle\sum\limits_{i=1}^3m_i^{3/2}}\right)\,\varepsilon_1^3.&
\end{eqnarray*}

By substitution of these and analogous relations into
Eqs.~(32), (42) and (43) we obtain the ultrarelativistic asymptotic
dependence for a state with an arbitrary type
$(n,k;n_1,n_2,n_3)$
\begin{equation}
J\simeq\alpha'E^2+\alpha_1E^{1/2},\qquad v_i\to1,\end{equation}
where
$$\alpha'=\frac1{2\pi\gamma}\,\frac n{n^2-k^2},\quad
\alpha_1=-\frac{\sqrt2\,n(n^2-k^2)^{-3/4}}{3\sqrt\pi\gamma}
{\textstyle\sum\limits_{i=1}^3m_i^{3/2}}\!.$$
That is close to the standard linear form $J=\alpha'E^2+\alpha_0$.

The slope coefficient in Eq.~(44) differs from the Nambu value for
the mesonic model $\alpha'=1/(2\pi\gamma)$ by the factor $n/(n^2-k^2)$.
This factor equals $3/8$ for the simple motions and attains the maximal
value $1/2$ (under admissible $n$ and $k$) for the ``quark-diquark"
motions with $n=2$, $k=0$. The latter case differs from the
quark-diquark baryon model only by the substitution
$\gamma\;\to\;2\gamma$.

The ``quark-diquark" state is preferable, if we assume
the principle of minimal energy --- the string system with given $J$
chooses the configuration with the minimal energy [7,8].

The first summand in Eq.~(42) that could be interpreted
as the ``string energy" or ``gluon energy" in the limit $v_i\to1$ or
$\varepsilon_i\to0$ grows as $\varepsilon_i^{-2}$, but the last
summands --- ``quark kinetic energy" $\sum m_i/\varepsilon_i$
grow as $\varepsilon_i^{-1}$. So in the
ultrarelativistic limit the ``string energy" dominates, and
the slope coefficient $\alpha'$ in (44) doesn't depend on
quark masses $m_i$.

The coefficient $\alpha_1$, otherwise, is determined by the
combination $\sum m_i^{3/2}$. This fact gives possibility
of estimating (in the model frameworks) the mentioned sum and
the quark masses $m_i$. This estimation will be accurate only for
baryons which satisfy two conditions: the quark motion is to be
relativistic and close to classic (the model is classic with
spinless quarks). The latter condition is equivalent to the standard
inequality $J/\hbar\gg1$ and in particular, results from the comparison
of typical sizes of the ``triangle" system in the
relativistic case (if $E\gg m_i\sqrt{1-v_i^2}$ in Eq.~(43))
$$R_i=\frac a\omega v_i\simeq\frac{2J}Ev_i\simeq3.95\cdot10^{-14}
\frac{J}\hbar\frac{\mbox{1 GeV}}E\frac{v_i}c \mbox{ cm}$$
with the corresponding length $\hbar/p\simeq\hbar/E$.
Furthermore, the motion is relativistic if the quarks are not very
heavy: $m_i\ll M=E$.

Express the combination $\sum m_i^{3/2}$ from Eq.~(44)
\begin{equation}
\sum_{i=1}^3m_i^{3/2}\simeq\frac{3(n^2-k^2)^{-1/4}}
{2^{3/2}\sqrt\pi}\bigg(E^{3/2}-\frac J{\alpha'\sqrt E}\bigg).
\end{equation}
It is natural to suppose that the states of the model are simple
($n=3$, $k=1$) or ``quark-diquark" ($n=2$, $k=0$). For these two
cases the mass estimations differ from each other by the small
factor $\simeq2^{1/6}$.

The expression in the parentheses in r.h.s. of Eq.~(45) is a small
difference of two large values. So it is very sensitive to errors in
$J$ and $E$. The simplest quantum correction to these values due to
quark spins implies an addition $S=\sum_{i=1}^3s_i$
(quark spin projection) to the classic angular momentum (43)
and $\Delta E=\Delta E_{SS}+\Delta E_{SO}$ --- to the energy (42).
The latter correction results from spin-spin ($\Delta E_{SS}$) and
spin-orbit ($\Delta E_{SO}$) interaction of quarks. The value
$\Delta E_{SO}$ is supposed to be due to pure Thomas precession
of quark spins [8,17], but there are some doubts as to the form and
the sign of this correction.
A precise form of $\Delta E$ is to be found only from a
consecutive quantum theory of this baryon model that isn't
constructed yet.

In the examples below the spin correction wasn't made.
The results of using Eq.~(45) for estimating quark masses
on examples of two Regge trajectories (nucleonic and for strange
$\Lambda$-particles) are shown in the following table. Masses of
$u$ and $d$-quarks are assumed to be equal. Here $m_{ud}$ and $m_s$
are effective quark masses measured in GeV; $J$ --- in units $\hbar$.
\begin{center}
\begin{tabular}{|c|c|c|c|}  \hline
$\quad J$& 1/2& 5/2& 9/2\\ \hline
{\scriptsize Particle}& $n,\,p$ &$N(1680)$&$N(2220)$\\ \hline
$m_{ud}$& $0.138\pm0.015$& $0.105\pm0.03$& $0.11\pm0.02$\\
\hline\hline
{\scriptsize Particle}&$\Lambda$&$\Lambda(1815)$&$\Lambda(2350)$\\
 \hline
$m_s$& $0.41\pm0.03$& $0.345\pm0.07$& $0.35\pm0.055$\\ \hline
$m_s^*$& $0.34\pm0.035$& $0.26\pm0.07$& $0.27\pm0.06$\\ \hline
\end{tabular}
\end{center}

The values $m_s$ were calculated under the assumption that
$m_{ud}\ll m_s$, and $m_s^*$ --- under the assumption
$m_{ud}\simeq0.1$ GeV.

The error ranges for $m_i$ are due to error
ranges in particle masses $M$ which influence the value
$\alpha'$.
The small difference between the results for the simple
and ``quark-diquark" configurations is also included in the
error ranges.

Note that the considered model (and other mentioned string models)
is applicable only to the orbitally excited baryon states (resonances)
with $J\ge5/2$ and isn't adequate for p, n and $\Lambda$-particles.

We may conclude that
calculated by Eq.~(45) quark masses are steady with growing $J$.
But the found values $m_{ud}\simeq100$ MeV and $m_s\simeq250$ MeV
(larger then other data for free quark masses [18] and less then
the constituent masses [8]) are preliminary and depend on the
spin correction.

The necessity of the spin correction is demonstrated by the following
fact. For the Regge trajectory with $\Delta$-resonances ($S=3/2$)
Eq.~(45) results in small negative values of $\sum m_i^{3/2}$
(error boxes include some positive range). But with substituting
$J -1/2$ instead of $J$ the formula (45) gives the steady value
$m_{ud}\simeq0.1$ GeV for heavy $\Delta$-resonances.

With growing $E$ and $J$ the influence of the unknown
$\Delta E$ on the values $m_i$ in Eq.~(45) diminishes, but too
slowly --- as $E^{-1/2}$ or $J^{-1/4}$. So for the available
baryon mass range 1 -- 3 GeV the spin correction in Eq.~(45) is
required for valid estimating the quark masses in the
frameworks of the considered model.

\bigskip

\centerline{\bf CONCLUSION}

\medskip

In the present paper a set of rotational motions of the baryon
model ``triangle" is investigated on the classic level. Quantization
in this model as in the string model of meson with massive ends [2,12]
encounters some problems connected with non-linear form of the
boundary conditions (8). Progress in this direction, for example,
description of quark spins will give possibility of precise model
prediction of the effective quark masses through comparison of calculated
dependence $J(E^2)$ with the experimental Regge trajectories.
But the problem of quantization needs special research which is
beyond the present paper.

On the other hand, the slope $\alpha'$ is finally determined by
Eq.~(44) on the classic level. It was mentioned that this coefficient
in the baryonic model ``triangle" differs from the mesonic slope
$\alpha'=1/(2\pi\gamma)$ by the factor $1/2$ for the ``quark-diquark"
states and by the factor $3/8$ for the simple states. The experimental
value $\alpha'\simeq1$\,GeV${}^{-2}$ is approximately equal for mesons
and baryons. So an effective value of string tension $\gamma$ in the
model ``triangle" is to be about $1/2$ or $3/8$ of the tension
in the model of meson. It is probably connected with different
energies of QCD interaction in the pairs: quark-quark and
quark-antiquark.

For the sake of comparison note that in the three-string model
[4\,--\,7] this factor equals $2/3$, i.e., the Regge slope in the
ultrarelativistic limit is $\alpha'=\frac23(2\pi\gamma)^{-1}$,
and the effective string tension is to differ by the same factor
from the mesonic one.

In the quark-diquark model and in the linear
($\cdot\!-\!\cdot\!-\!\cdot$) configuration the Regge slope
$\alpha'=1/(2\pi\gamma)$ equals the mesonic one. So these
models explain the equality of values $\alpha'$ for mesons and
baryons by the natural way (the rotational motions of these models
are meson-like). But this advantage is balanced an explicit
dissymetry of the quarks in the both models. Furthermore, the
``triangle" and Y configurations unlike two others string baryon
models are QSD-motivated in the Wilson loop operator approach [10].

For description of baryons on the Regge trajectories the
``quark-diquark" states and the simple states (Fig. 2)
of the model ``triangle" were used. Under the assumption that
the energy of the orbitally excited string state for the
given angular momentum $J$ is minimal [7,8] these configurations
are preferable.

The exotic states naturally emerging in this model
are probably too exotic for applications in particle physics,
(except for hybrids).

\medskip

\centerline{\bf Acknowledgment}

\medskip

The author is deeply indebted to Drs.  B.M.\,Barbashov and
V.V.\,Nesterenko for the useful discussion.

\medskip

\centerline{\bf References}

\medskip

\noindent[1] A. Chodos and C.B. Thorn, Nucl. phys. {\bf B72}, 509
(1974).
\newline[2] B.M. Barbashov and V.V. Nesterenko, Introduction to the
relativistic string theory. Singapore: World scientific, 1990.
\newline[3] B.M. Barbashov, In Proc. of the Conf. ``Strong Interactions at
Long Distances" / Ed. L. Jenkovsky, Hadronic
Press, Palm Harbor, FL, USA, ISBN 0-911767-99-1, P. 257--275 (1995).
\newline[4] X. Artru, Nucl. phys. {\bf B85}, 442 (1975);
Phys. Rep. {\bf 97}, 147 (1983).
\newline[5] P.A. Collins and J.F.L. Hopkinson, ibid. {\bf B100}, 157
(1975); K. Sundermeyer and A. de la Torre, Phys.Rev.D {\bf15},
1745 (1977).
\newline[6] M.S. Plyuschai, G.P. Pron'ko and A.V. Razumov,
Theor. Mathem. Physics. {\bf 63}, 389 (1985); S.V. Klimenko et al.,
Theor. Mathem. Physics. {\bf 64}, 810 (1985).
\newline[7] K. Johnson and C.B. Thorn, Phys. Rev. D {\bf 13}, 1934 (1976).
\newline[8] Yu.I. Kobzarev, L.A. Kondratyuk, B.V. Martemyanov
and M.G. Schepkin, Yad. Phys. {\bf 45}, 526 (1987).
\newline[9] G.S. Sharov, Theor. Mathem. Physics. {\bf 113}, 1263 (1997).
\newline[10] John M. Cornwall, Phys. Rev. D {\bf 54}, 6527 (1996).
\newline[11] Dan LaCourse and M.C. Olsson, Phys. Rev. D {\bf 39}, 2751
(1989); M.C. Olsson and Ken Williams, Phys. Rev. D {\bf 48}, 417 (1993).
\newline[12] G.S. Sharov, Theor. Mathem. Physics. {\bf 107}, 487 (1996).
\newline[13] B.M. Barbashov and G.S. Sharov, Theor. Mathem. Physics.
{\bf 101}, 1332 (1994).
\newline[14] J.F. Traub, Iterative methods for the solutions of
equations. Chelsea Publishing Company. New York, N.Y., 1982.
\newline[15] D. Hilbert, S. Cohn-Vossen, Anschauliche Geometrie.
Berlin, 1932.
\newline[16] V.P. Petrov and G.S. Sharov, Theor. Mathem. Physics.
{\bf 109}, 1388 (1996).
\newline[17] T.J. Allen, M.C. Olsson, S. Veseli and Ken Williams,
 Phys. Rev. D {\bf 55}, 5408 (1997).
\newline[18]  Particle Data Group: R.M. Barnett et al. Phys. Rev. D {\bf 54}, No. 1, Part I (1996).
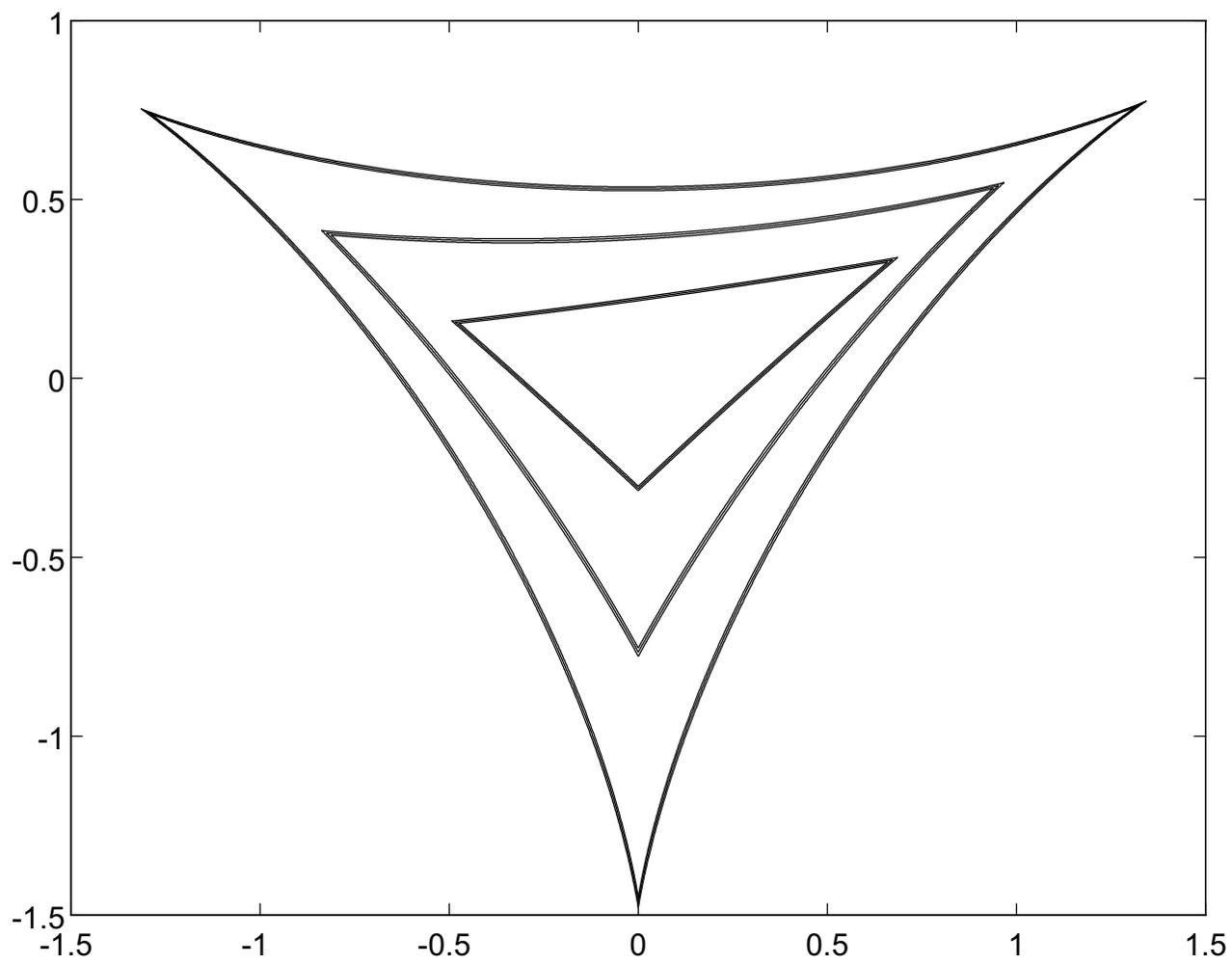
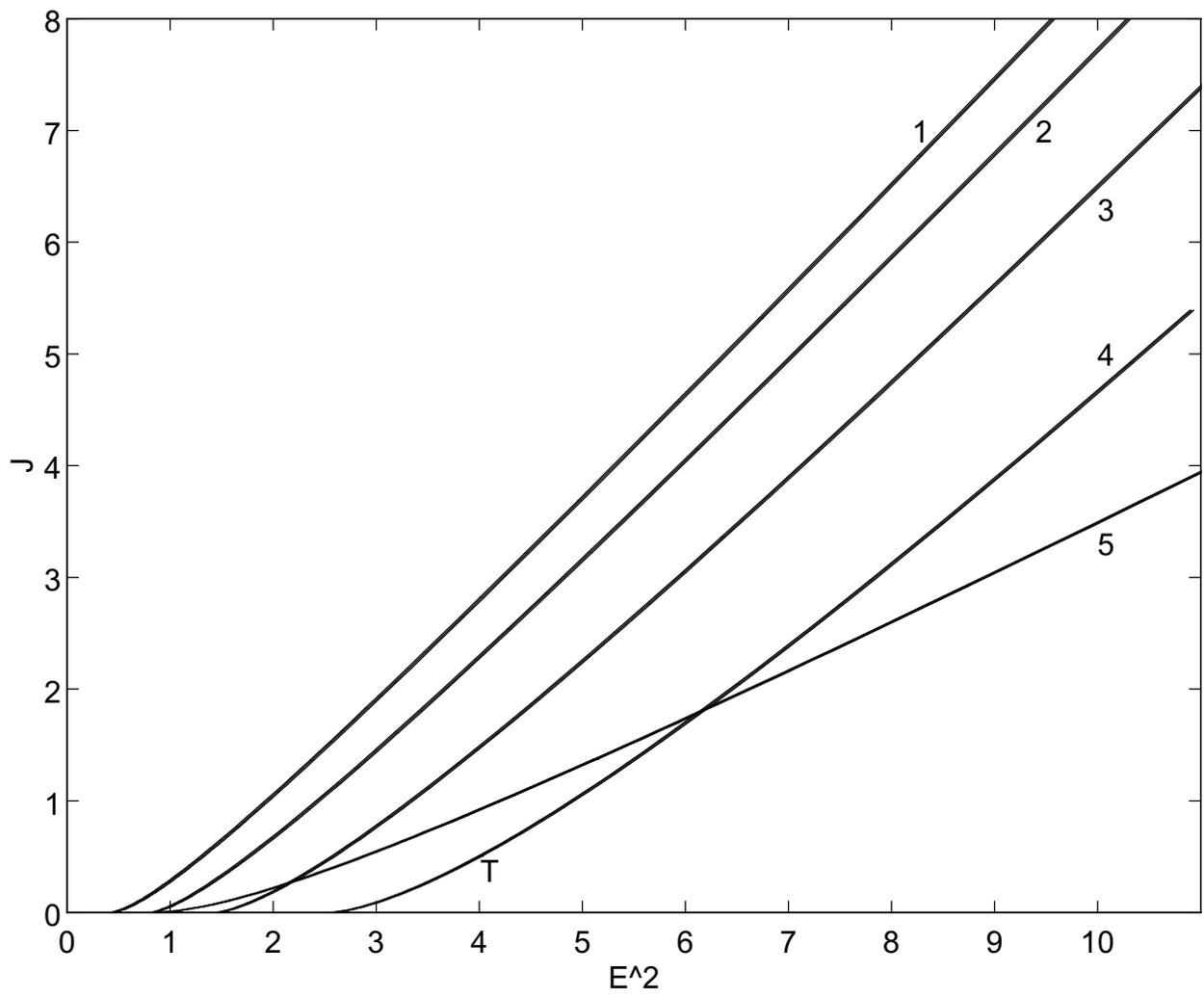
\end{document}